\documentclass[12pt]{article}

\usepackage{amssymb}
\usepackage{graphicx}% Include figure files
\usepackage{subfigure}
\usepackage{hyperref}% add hypertext capabilities
\usepackage{mathrsfs}
\usepackage{latexsym}
\usepackage{amsfonts}
\usepackage{amsmath}
\usepackage{esint}
\usepackage{amsxtra}
\usepackage[enableskew]{youngtab}

\newdimen\mathindent
\mathindent\leftmargini

\def \ep1{\epsilon_1}
\def \ep2{\epsilon_2}

\def \be{\begin{equation}}
\def \ee{\end{equation}}
\def\beq{\begin{eqnarray}}
\def\eeq{\end{eqnarray}}
\def \ba{\begin{array}}
\def \ea{\end{array}}
\def\nn{\nonumber}

\def \f{\frac}

\def \p{\partial}

\def \ep{\epsilon}

\begin{document}
\vspace*{-.6in}
\thispagestyle{empty}
\baselineskip = 18pt

\vspace{.5in}
\vspace{.5in}

{\LARGE
\begin{center}
Quasimodular instanton partition function and\\ the elliptic solution of Korteweg-de Vries equations
\end{center}}

\vspace{1.0cm}

\begin{center}
Wei He\footnote{weihephys@gmail.com}\\
\vspace{1.0cm}

\em{Instituto de F\'isica Te\'orica, Universidade Estadual Paulista,\\
Barra Funda, 01140-070, S\~{a}o Paulo, SP, Brazil}

\end{center}
\vspace{1.0cm}

\begin{center}
\textbf{Abstract}
\end{center}
\begin{quotation}
\noindent
The Gauge/Bethe correspondence relates Omega-deformed N=2 supersymmetric gauge theories to some quantum integrable models, in simple cases the integrable models can be treated as solvable quantum mechanics models.
For SU(2) gauge theory with an adjoint matter, or with 4 fundamental matters, the potential of corresponding quantum model is
the elliptic function. If the mass of matter takes special value then the potential is an elliptic solution of KdV hierarchy.
We show that the deformed prepotential of gauge theory can be obtained from the average
densities of conserved charges of the classical KdV solution, the UV
gauge coupling dependence is assembled into the Eisenstein series.
The gauge theory with adjoint mass is taken as the example.
\end{quotation}

\section{Introduction}

It is well-known that the Korteweg-de Vries equation(KdV) is an
integrable Hamiltonian system of infinite dimension \cite{zf1971}.
The rapidly decreasing solutions are obtained by the inverse scattering method \cite{ggkm1967}, where the
scattering data of the Schr\"{o}dinger operator with the initial
value KdV function $u(x,t=0)$ as the potential can be used to
reconstruct the exact potential. In the case of reflectionless
scattering, the rapidly decreasing soliton solutions are obtained.

The analogous problem of solving KdV equations with periodic initial
condition leads to the discovery of relations to some other topics,
including the elliptic functions, algebraic geometric methods and the
finite gap spectrum problem \cite{novikov, lax}. The inverse problem for
the periodic potential is related to the spectral problem of Schr\"{o}dinger operator with Dirichlet boundary
condition. The periodic potentials are sometimes called periodic
KdV soliton, albeit many aspects of the solution do not parallel
with the fast decaying solitons.

The KdV system is a classical Hamiltonian system, a question we can
ask is how to quantize the KdV system, and especially how to
quantize the soliton solution? For the first half of the question,
it was noticed that the Poisson bracket of the KdV hierarchy is the
large central charge limit of the Virasoro algebra
\cite{Gervais1985}, therefore the conformal field theory (CFT)
provides a well defined framework to deform the classical KdV system
where the inverse of central charge is the deformation parameter. In
the literature, the deformation procedure based on CFT is often
called ``quantization" because the procedure is similar to quantum
mechanics deforming classical mechanics. In fact, the quantum
Hamiltonians of some soliton equations, including the closely
related sine-Gordon, KdV, and mKdV equations, can be constructed
from CFT energy momentum tensor and its derivatives, they are in
involution with respect to the Virasoro algebra, and quantum soliton
equations can be formally defined in this context \cite{sy1988,
ey1989, blz1}. These discoveries finally cumulate to a series of
papers studying the integrable structure of CFT, starting from
\cite{blz1}. The second half of the question is also well motivated.
The quantum behavior of soliton shows some remarkable properties
such as particle-soliton duality, which are absent at the classical
level. In 1+1 dimension we have the example of the
sine-Gordon/Thirring model duality \cite{coleman1975}, in higher
dimension we have the Montonen-Olive duality for gauge
theories \cite{mo1977}.

In this paper we are interested in ``quantization" of the periodic
solutions mentioned above. However, there is not a first principle
method to identify the corresponding quantity in CFT. We have
a hint from the linear equation with periodic potential associated to KdV equation. This equation can be viewed
from two directions. Firstly, according to the Gauge/Bethe
correspondence such equation is a  Schr\"{o}dinger  equation for some quantum mechanical models which are related to the $\Omega$-deformed 4-dimensional N=2 supersymmetric Yang-Mills gauge theories in the Nekrasov-Shatashvili
(NS) limit $\ep_1\ne0,\ep_2=0$ \cite{ns}. On the other hand, the linear
equation is a classical limit of the null operator decoupling equation of certain
Liouville CFT correlation function on surfaces, see e.g. \cite{mt1006}.
Conformal blocks of Liouville CFT are related to Nekrasov partition functions of the mentioned above $\Omega$-deformed N=2
supersymmetric gauge theories, according to the
Alday-Gaiotto-Tachikawa (AGT) correspondence \cite{agt}. Therefore,
with these relations we can relate some periodic KdV solutions to
certain CFT on particular surfaces, and also related them to
some deformed N=2 quantum gauge theory models.

We use a particular example to examine in detail the relations stated above.
The periodic KdV solution is the Lam\'{e} potential which is a
prototype of elliptic KdV soliton \cite{novikov, lax}, the linear
equation describes quantum particle in the elliptic potential
\cite{ns}, the associated equation of CFT is the null vector decoupling equation for the 2-point function on the
torus \cite{agt}, and the gauge theory is the $N=2^*$ theory
\cite{SW9407, SW9408, instcount}.

The results of the paper can be divided into two parts. In the first
part, we first compute the eigenvalue function of the
Schr\"{o}dinger equation from classical KdV Hamiltonians of elliptic
solution (\ref{classicalcharge}), this method is different from the
commonly used WKB perturbation. Then according to the Gauge/Bethe
correspondence, the spectral solution of the Schr\"{o}dinger problem
gives the twisted superpotential of $\ep_1$ deformed gauge theory,
which is the prepotential of $\Omega$-deformed gauge theory in the NS limit. Therefore we obtain the $\ep_1$ deformed prepotential
from classical KdV Hamiltonians, with all $q$ dependence contained
in polynomials of Eisenstein series, in formula (\ref{potenNS}). In
the second part we study the correspondence without taking the limit
$\ep_2\to0$. The computation is based on the Nekrasov partition
function for generic parameter $\ep_1, \ep_2$, we propose a
generalized spectral relation given in (\ref{quantumlambda}), then
reverse the relation we get a deformed version of KdV charges
(\ref{quantumcharge}). As the leading order of them are the
classical charges, and they contain subleading corrections, therefore
we interpret them as the quantum charges of elliptic KdV solution.
Similar to the $\ep_1$ deformed case in the first part, here we can
write $\ep_1, \ep_2$ deformed prepotential in a form that the
coefficients are quasimodular forms, presented in
(\ref{modulpoten0}). The quasimodular expansion of prepotential has
been studied by other methods, including solving the
holomorphic/modular anomaly equation, for the undeformed Seiberg-Witten
theory in \cite{mnw9710}, for the deformed theory in \cite{hkk, bfglp,
bfglp2}, and performing the WKB perturbation for the elliptic
potential when $\ep_2=0$ \cite{kpt, piatek}. Our approach gives a new perspective on its origin.

There is another elliptic potential, the Treibich-Verdier potential,
which can be studied in the same way. At the moment it is not clear
to us if the connections can be generalized to other KdV solutions,
this is remained for future work.

The organization of the paper is as follows. In Section 2, we
compute the average densities of conserved charges for the Lam\'{e}
elliptic potential, from which we obtain the quasimodular form of
prepotential of $N=2^*$ SYM in the Omega background in NS limit. In
Section 3 we briefly explain the connection of conformal field
theory and quantization of KdV system. In Section 4, based on the
AGT connection we use CFT/gauge theory technique to compute the
spectrum of elliptic solution which is naturally interpreted as
quantum spectrum of the solution. Section 5 is devoted to some open
questions.

\section{Classical spectrum of elliptic solution of KdV equations}

\subsection{From gauge theory to Schr\"{o}dinger operator and KdV}

Our story starts from the Gauge/Bethe correspondence which relates
the Coulomb vacuum of N=2 supersymmetric gauge theories with
$(\ep_1=\ep,\ep_2=0)$ deformation to the Bethe solution of some
quantum integrable models  \cite{ns}. In some simple cases, the
SU(2) gauge theories are related to one particle quantum mechanical
models with periodic potentials, \be
(\p^2-u)\Psi(x)=\lambda\Psi(x),\qquad u(x)=u(x+T).\label{lineq}\ee
In these cases we have analytical results about both gauge theory
and quantum mechanics models, we can precisely test the idea of the
Gauge/Bethe correspondence. An important relation is about the
eigenvalue $\lambda$ of Schr\"{o}dinger operator and the deformed
prepotential $\mathcal{F}(\ep_1)$ of gauge theory. The precise
relation depends on models. The Floquet theorem indicates the
monodromy of the wave function along the period $T$ takes the form
$\Psi(x+T)=\exp(i\nu(\lambda)T)\Psi(x)$. A key point here is that the spectrum of Schr\"{o}dinger
operator is given by the inverse of the function $\nu(\lambda)$, therefore we can solve the spectral problem if we can
compute the function $\nu(\lambda)$.

We can follow conventional way to perform the WKB analysis for the
equation (\ref{lineq}) and obtain the function $\nu(\lambda)$
without knowing solution of wave function \cite{he1108}. However,
here we apply a different and simpler method to compute the function
$\nu(\lambda)$. The Schr\"{o}dinger operator is related to the KdV
hierarchy, the coefficients of large $\lambda$ asymptotic expansion
of the monodromy are KdV Hamiltonians, as we will explain
momentarily.

That is not the whole story, the linear equation (\ref{lineq}) is
further related to theory of integrable quantum field theory. In our
story the integrable field theory is Liouville CFT \cite{agt}. A
natural next step is to study what is the story for KdV theory if we
do not take limit $\ep_2\to0$ for the CFT/gauge theory. Indeed,
without taking classical limit for the CFT a quantum
KdV(qKdV) theory was developed in a series of papers, for example
\cite{sy1988, ey1989, blz1}. Therefore it seems that there is a
deformed version of all the relations mentioned above.

The results present in this section is a limit case of the qKdV-CFT-gauge theory connection, it is the classical limit of qKdV, which means the large central charge limit of CFT,  or the NS limit of gauge
theory. In the next two sections we discuss the quantum KdV solution, with proper interpretation,  where it appears that similar to the classical solution the quantum spectrum also displays quasimodular structure.

\subsection{Densities of conserved charges of elliptic solutions}

The linear operator $L=\p^2-u(x,t)$ plays a central role in the
theory of KdV hierarchy. In the Lax's operator formalism of KdV
hierarchy, a tower of differential operators with increasing order
would give the higher order KdV equations. These operators can be
obtained from the formal computation of pseudo-differential operator
system of Gelfand and Dickey, the equations appear as
$\p_{t_n}L=[(L^{(2n-1)/2})_+, L]$, with $ n=1,2,3,\cdots$. A basic
fact is that $L$ is isospectral for the KdV solution $u$, therefore
it is enough to consider the spectrum of $L=\p^2-u(x)$ with
$u(x)=u(x,0)$.

There is a nice way to see how the Hamiltonians and the monodromy
are related. We start from the linear system of $L$ in
(\ref{lineq}), let $\Psi(x)=\exp(\int^x v(y)dy)$ and substitute it
into equation (\ref{lineq}), we get the Miura transformation, \be
v^{'}+v^2=u+\lambda. \label{miura}\ee In accordance with literature,
we use $\p$ or $'$ to denote $\p_x$. Suppose the spectral parameter
$\lambda\gg 1$, perform the asymptotic expansion for $v(x)$, \be
v=\sqrt{\lambda}+\sum_{k=1}^{\infty}\f{v_k}{(\sqrt{\lambda})^k},\ee
then we obtain all $v_k$ as functionals of $u$ and its derivatives,
$v_k=v_k(u,u^{'},u^{''},\cdots)$, they can be determined recursively
by the Miura transformation. For fast decaying or periodic
potentials, as $v_{2k}$ are total derivatives, the nontrivial KdV
dynamics are driven by $v_{2k-1}$. The KdV Hamiltonians are defined
as integration of the densities $v_{2k-1}$, \be H_{k}=\int
dxv_{2k-1}.\ee The first few of them are \be
\begin{aligned}
H_1&=\f{1}{2}\int dxu,\qquad H_2=-\f{1}{8}\int dx(u^2-u^{''}),\\
H_3&=\f{1}{32}\int
dx(2u^3+u^{'2}+(u^{'''}-6uu^{'})^{'}),\qquad \cdots,
\end{aligned}
\ee
they are in involution with respect to the Poisson structure of KdV.
Restore the time dependence, the KdV equations are $\p_{t_n}u=\lbrace
H_n, u\rbrace$. The Miura transformation is related to the
bi-Hamiltonian structure of KdV hierarchy.

Now let us turn to the periodic solutions of KdV equations, the
initial profile $u(x, 0)$ is given by the solution of generalized
$n$th stationary equation $\lbrace \sum_{k=1}^{n}c_kH_k,
u\rbrace=0$, where $c_k$ are coefficients \cite{novikov}. We are interested in a
class of periodic solutions given by the Weierstrass elliptic
function, in the associated linear spectral problem they are called
finite gap potentials. One periodic solution is the elliptic
Lam\'{e} potential, \be u(x)=n(n-1)\wp(x;\omega_1,\omega_2),\qquad
n\in\mathbb{Z}_+,\ee it solves the $(n-1)$th stationary KdV
equation, and $n-1$ would be the number of gaps for the real
spectrum of $L$, the arithmetic genus of the associated surface. The
nome for the elliptic function is
$q=\exp(i2\pi\f{\omega_2}{\omega_1})$.

Recall that for the Schr\"{o}dinger operator $L$ with potential $u(x)$ with period $T$,
the monodromy of the wave function under $x\to x+T$ is given by $\int_y^{y+T}dxv(x)=i\nu T$, therefore the Floquet exponent is given by
\be i\nu=\f{1}{T}\int_y^{y+T} dxv(x).\label{monodPsi}\ee
For large $\lambda$, we can use the asymptotic expansion of $v(x)$.
The integration $\varepsilon_{k}=\f{1}{T}\int_y^{y+T}dxv_{2k-1}$ is the average density of Hamiltonian $H_k$, $H_{k}=T\varepsilon_{k}$. For elliptic function there are two periods, as we have examined the large $\lambda$ asymptotic expansion is associated to $T=2\omega_1$ \cite{he1108}. The other period is related to dual expansion which is not suitable for the KdV formalism.

Substitute the Lam\'{e} potential into the Hamiltonians, we can simplify all the integrands $v_{2k-1}(x)$ using the basic relations of the $\wp(x;\omega_1,\omega_2)$ function and discard total derivative terms.
The final form of the integrands includes two parts, one part is $x$-independent and the other part is proportional to $\wp(x)$.
The expansion of Floquet exponent takes the following form \be
i\nu=\sqrt{\lambda}+\sum_{k=1}^{\infty}\f{\varepsilon_{k}(n,g_{2,3},\f{\zeta_1}{\omega_1})}{(\sqrt{\lambda})^{2k-1}},\label{nulambda}\ee
where $g_2, g_3$ are invariants of the $\wp(x)$ function, the integration of $\wp(x)$
zeta function is $\zeta(x)$, satisfying $\p_x\zeta(x)=-\wp(x)$, and its periodic shift gives
$\zeta(x+2\omega_{1,2})=\zeta(x)+2\zeta_{1,2}$.
The average density $\varepsilon_k$ can be written in terms of Eisenstein series if we use the relation \be
\f{\zeta_1}{\omega_1}=\f{\pi^2}{3}E_2(q),\quad
g_2=\f{4\pi^4}{3}E_4(q),\quad g_3=\f{8\pi^6}{27}E_6(q).\ee
Under the SL(2, $\mathbb{Z}$) transformation $E_4, E_6$ are modular forms of weight 4 and 6, respectively, while $E_2$ is a quasimodular form.
The first few $\varepsilon_k$ are
\be
\begin{aligned}
\varepsilon_1=&-\f{\pi^2}{6}n(n-1)E_2,\\
\varepsilon_2=&-\f{\pi^4}{72}n^2(n-1)^2E_4,\\
\varepsilon_3=&-\f{\pi^6}{2160}n^3(n-1)^3(9E_2E_4-4E_6)+\f{\pi^6}{180}n^2(n-1)^2(E_2E_4-E_6),\\
\varepsilon_4=&-\f{5\pi^8}{72576}n^4(n-1)^4(15E_4^2-8E_2E_6)+\f{5\pi^8}{1512}n^3(n-1)^3(E_4^2-E_2E_6)\\
&-\f{\pi^8}{252}n^2(n-1)^2(E_4^2-E_2E_6),\\
\cdots\label{classicalcharge}
\end{aligned}
\ee
Compare to the WKB computation in \cite{he1108}, where we essentially performed the $\ep$-perturbation expansion, the KdV method we employ is based on the large eigenvalue expansion. The integrand from the KdV densities are polynomials of $u(x)$ and its derivatives, the computation is much simpler.

Now, we can use the Bethe/Gauge correspondence to derive the deformed prepotential of gauge theory from the spectral data of Schr\"{o}dinger equation \cite{ns}.
In the next subsection we will show that from the spectrum of classical elliptic solution we exactly recover the deformed instanton action of SU(2) $N=2^*$ gauge theory.

\subsection{Instanton partition function in the NS limit}

We already obtained the asymptotic relation $\nu(\lambda)$, the
reverse relation gives us the asymptotic spectrum of the Lam\'{e}
potential. Then according the proposal in \cite{ns}, N=2 gauge
theories in the limit $\ep_1=\ep,\ep_2=0$ are related to quantum
integrable models. In our story, Schr\"{o}dinger equation with the
Lam\'{e} potential is related to the gauge theory is the mass
deformed SU(2) N=4 super-Yang-Mills theory, i.e.  $N=2^*$ theory. The gauge theory mass
appears in the potential as $m(m-\ep_1)\wp(x)$. Because we relate the gauge theory to the KdV solution we demand the
mass of adjoint matter takes special value so the potential satisfies a stationary KdV equation,\be
\f{m}{\epsilon_1}=n\in\mathbb{Z}.\ee As checked in detail in
\cite{he1108}, the spectral data $\lambda, \nu$ are related to gauge
theory quantities, the moduli $\tilde{\mathfrak{u}}$ and the v.e.v.
of scalar field $a$, by relation \footnote{ Compare to \cite{he1108} here we
have recovered the $\pi$ factors.} \be
\lambda=-\f{8\pi^2\tilde{\mathfrak{u}}}{\epsilon_1^2},\qquad
\nu=\f{2\pi a}{\epsilon_1},\ee and
$\tilde{\mathfrak{u}}=\tilde{\mathfrak{u}}(a,m,q,\epsilon_1)$ is
related to the deformed prepotential of gauge theory in the NS limit
$\mathcal{F}(a, m, q, \ep_1)$ by \be
\tilde{\mathfrak{u}}=\f{1}{2}q\f{\p}{\p
q}\mathcal{F}(\ep_1)+\f{m(m-\ep_1)}{24}(1-2E_2).\label{ufrompoten}\ee
Here the same $q$ is the complex UV coupling of gauge theory, the
instanton expansion parameter. The second term on the right hand
side of (\ref{ufrompoten}) is necessary to march the eigenvalue
$\lambda$ and the gauge theory prepotential $\mathcal{F}(\ep_1)$. We
can derive the  Schr\"{o}dinger equation by examining $\ep_2\to0$
limit of the null vector decoupling equation of Liouville CFT, treating chiral half of
the correlation function as the wave function, the relation
(\ref{ufrompoten}) can be derived \cite{piatek}.

In order to relate the spectrum of KdV solution to the instanton
partition function, we need to reverse the relation
(\ref{nulambda}), and get the large $\nu$-expansion for the spectral parameter,
\be
\lambda=-\nu^2+\sum_{k=0}^{\infty}\f{\lambda_{k}^c}{\nu^{2k}},\label{lambdanu}\ee
where $\lambda_0^c=-2\varepsilon_1, \lambda_1^c=\varepsilon_1^2+2\varepsilon_2, \lambda_2^c=-2(\varepsilon_1^3+3\varepsilon_1\varepsilon_2+\varepsilon_3),\cdots$. Substitute $\varepsilon_k$, we obtain
\be
\begin{aligned}
\lambda_0^c=&\f{\pi^2}{3}n(n-1)E_2,\\
\lambda_1^c=&\f{\pi^4}{36}n^2(n-1)^2(E_2^2-E_4),\\
\lambda_2^c=&\f{\pi^6}{540}n^3(n-1)^3(5E_2^3-3E_2E_4-2E_6)-\f{\pi^6}{90}n^2(n-1)^2(E_2E_4-E_6),\\
\lambda_3^c=&\f{\pi^8}{9072}n^4(n-1)^4(35E_2^4-7E_2^2E_4-10E_4^2-18E_2E_6)-\f{\pi^8}{756}n^3(n-1)^3(7E_2^2E_4-5E_4^2-2E_2E_6)\\
&+\f{\pi^8}{126}n^2(n-1)^2(E_2E_6-E_4^2),\\
\cdots\label{classicallambda}
\end{aligned}
\ee  This is the asymptotic spectrum of the linear Schr\"{o}dinger
operator $L$ \cite{he1108}, but now wrote in a different form: the
$\ep_1$ is hidden in $n$, the $q$ is assemble into the quasimodular
functions $E_{2k}$.

Using the differential relation of the (quasi)modular functions, \be
\begin{aligned}
&q\f{\p}{\p q}\ln\eta(q)=\f{1}{24}E_2,\qquad
q\f{\p}{\p q}E_2=\f{1}{12}(E_2^2-E_4),\\
&q\f{\p}{\p q}E_4=\f{1}{3}(E_2E_4-E_6),\qquad q\f{\p}{\p
q}E_6=\f{1}{2}(E_2E_6-E_4^2),\label{difE}
\end{aligned}
\ee  where $\eta(q)$ is the Dedekind eta function, we can integrate
(\ref{ufrompoten}) to recover the gauge theory prepotential up to a $q$-independent integration constant, \be
\begin{aligned}
\widetilde{\mathcal{F}}(\ep_1)=&(a^2-\f{m(m-\ep_1)}{12})\ln q+2m(m-\ep_1)\ln\eta(q)-\f{m^2(m-\ep_1)^2}{48a^2}E_2\\
&-\f{1}{5760a^4}[m^3(m-\ep_1)^3(5E_2^2+
E_4)-\ep_1^2m^2(m-\ep_1)^2E_4]+\mathcal{O}(a^{-5})\label{potenNS}
\end{aligned}
\ee
Note that $\ln\eta(q)=\f{1}{24}\ln q-q+\mathcal{O}(q^2)$, therefore terms of order $m(m-\ep_1)\ln q$ cancel out.

There are two subtle points about the above formula (\ref{potenNS}) concerning $q$-independent terms.
Firstly, the integration of the last three relations in (\ref{difE})
about modular forms introduces some $q$-independent constant terms of definitive value.
This constant part is just the $q=0$ value of terms involving
$E_{2k}$ in (\ref{potenNS}), it should not be contained in the
instanton part of deformed prepotential. However, we do not need to subtract
it from the prepotential (\ref{potenNS}) because it precisely gives
most terms of the 1-loop perturbative part of the deformed prepotential.
Secondly, there is a remaining 1-loop perturbative $q$-independent piece of order $m^2\ln2a$ which cannot
be determined by our integration procedure. Together with $\widetilde{\mathcal{F}}(\ep_1)$, they give the full deformed prepotential in the NS limit,
\be
\mathcal{F}(\ep_1)=\widetilde{\mathcal{F}}(\ep_1)+m(m-\ep_1)\ln2a.\ee
Therefore we recover {\em almost} the full prepotential in the NS limit by integrating the classical elliptic KdV solution.
In the Section 4 we will discuss these points in more detail when both deformation parameters
$\ep_1,\ep_2$ are turned on where the 1-loop deformed prepotential is computed by the double gamma function.

In the limit $\ep_1\to0$, in the gauge theory we get the solution
of undeformed prepotential, see the results in \cite{mnw9710} computed from the Seiberg-Witten curve.
In quantum mechanics this limit is the leading order WKB approximation, see e.g. \cite{piatek}. In KdV theory this
limit is the dispersionless limit, under which only one term in
$v_{2k-1}$ survives and the Hamiltonians are very simple,
$H_k\propto \int dxu^k$.

\section{Quantum KdV hierarchy and CFT}

Now we keep both deformation parameters $\ep_1,\ep_2$, and look at how the $\ep_2$ deforms all the classical relations discussed previously. The full $\ep_1,\ep_2$ deformation of gauge theory \cite{instcount}, and its relation to CFT are already understood \cite{agt}, we would study how the corresponding elliptic solution of KdV hierarchy is deformed.  As we explain below, for elliptic KdV solution it is difficult to follow the already known examples to construct quantum solution which produces the elliptic solution in the classical limit.

Let us briefly recall how the KdV theory is related to CFT.
We have no a prior quantization scheme in KdV theory because it is a nonlinear system without adjustable couplings.
The clue is the symmetry algebra,  there are two Poisson structures for the KdV system, one of the Poisson brackets,
given by the Magri-Virasoro bracket, is just the large central
charge limit of Virasoro commutation relation \cite{Gervais1985}.
Therefore we can map the potential $u(x)$ to the energy-momentum operator $T(z)$ in CFT, and continue to construct operators that are commutative with $T(z)$, they are operator analogues of the Hamiltonian $H_k$.
Consider the CFT in the complex plane, or conformal equivalently on the infinite cylinder,
the quantum actions are defined
through the normal ordered product of CFT energy-momentum operator and its
derivatives \cite{sy1988, ey1989, blz1},
\be
\begin{aligned}
I_1&=\int dzT(z),\qquad I_2=\int dz:T^2(z):,\\
I_3&=\int dz:T^3(z):+\f{c+2}{12}:(T^{'}(z))^2:,\qquad\cdots
\label{qkdv1}\end{aligned} \ee they commutate with respect to the
Virasoro algebra, and the quantum KdV equations can be formally
defined. After explicitly carrying out integration the quantum Hamiltonians
$I_n$ are polynomials of Virasoro generators. In the classical
limit, $c\to\infty$, \be T(z)\to\f{c}{6}u(z),\qquad
[*,*]\to\f{6i\pi}{c}\lbrace*,*\rbrace,\label{qkdv2}\ee we recover
classical KdV theory.

Many ingredients of the classical theory have a corresponding quantum
version. In classical theory the Miura transformation (\ref{miura})
maps two Poisson structures satisfied by $u(x)$ and $v(x)$
respectively to each other, in the quantum version it is the Feigin-Fuchs free
field realization of CFT. The ordinary differential equation of the
classical theory is replaced by the null vector decoupling condition of CFT which is a partial differential equation.

This construction clarifies the connection of classical and quantum
symmetry algebra, but it has not yet addressed some other points of KdV
theory. The soliton solution of classical KdV equations is an
important aspect, study the quantum version of KdV solitons may
provide further insight about them. Generally, a CFT state in the Verma module,
measured by its conformal weight,  would be a solution to the
quantum KdV system. In order to survive in the large $c$ limit, the
conformal weight should be of order $\mathcal{O}(c)$. In the full
quantum theory, the classical profile may be dressed up with light
excitations created by $L_{-n}$ with conformal weight of order
$\mathcal{O}(1)$. At the moment it is not clear in general what kind
of CFT states, in the classical limit, would give various classical
solutions of KdV equations.

The CFT dynamics on the cylinder can be directly mapped to
the classical KdV dynamics on the circle, as in
(\ref{qkdv1}),(\ref{qkdv2}). A novel feature of 2-dimensional CFT is
that they can be defined on Riemann surface of nontrivial topology,
i.e. surface of higher genus with punctures and boundaries. What is
the corresponding story for KdV theory when the CFT is on surfaces
of nontrivial topology? There is no obvious general answer for this
question. In the case of complex plane the eigenstates for the quantum Hamiltonians can be defined by the radial quantization,
then eigenvalues can be computed \cite{blz1}. However, in the case of nontrivial surfaces or complex plane with branch cuts,
it is unclear how to define eigenstates in CFT.

Indeed, some elliptic solutions are related to CFT on nontrivial surfaces.
The classical limit of null vector decoupling equation on the torus gives the Schr\"{o}dinger equation with the Lam\'{e} potential,
the null vector decoupling equation on the sphere leads to the Schr\"{o}dinger equation with the Treibich-Verdier potential, see e.g. \cite{mt1006, he1306, piatek}.
The wave function is the classical limit of the chiral half of the correlator with a degenerate operator insertion.
This fact gives us some hint about the deformed story.
We computed classical KdV charges from the monodromy of wave function, we can also try to compute the deformed charges from the monodromy of conformal block when moving the degenerate operator along the conjugate cycles on the surface,
which corresponds to the expectation value of the Wilson loop in gauge theory \cite{loop1, loop2}.
For example, for the torus conformal block the $\alpha$-cycle monodromy of  the degenerate operator is a simple phase $\exp(i\f{2\pi a}{\ep_1})$,
in the classical limit it is the same as the monodromy of Schr\"{o}dinger wave function (\ref{monodPsi}) whose expansion (\ref{nulambda}) gives the classical KdV charges.

We should express the deformed quantities as classical part plus large $c$ corrections.
The CFT, or equivalently the N=2 gauge theory as \cite{agt}
indicates, provides effective tools to compute the large central charge
deformation, therefore we can interpret the subleading terms as
quantum effects for the periodic solutions. The $\Omega$ background
parameters of gauge theory are related to the central charge of CFT
by\be c=1+6\f{(\ep_1+\ep_2)^2}{\ep_1\ep_2}.\ee Therefore the NS
limit $\ep_1=\ep,\ep_2\to0$ in gauge theory is the classical limit
of CFT, $c\to\infty$. In the case of torus 1-point block, the
external state has a conformal weight \be
\Delta_m=\f{m(\ep_1+\ep_2-m)}{\ep_1\ep_2}.\ee In the classical limit
$\Delta_m\sim\mathcal{O}(n(n-1)c)$, hence it produces the Lam\'{e}
potential. Instead of performing large $c$ expansion, we define a
deformation parameter by \be \hbar=\f{\ep_2}{\ep_1}.\ee It can be
views as the quantum parameter for the nonlinear KdV system because
we assume $\hbar$ is small. $\ep_1$ is the quantum parameter for the
associated linear equation (\ref{lineq}), here it is used to make
the coefficient of elliptic potential to be triangular numbers
$n(n-1)$ in order to satisfy the stationary KdV equations.

\section{Quasimodular instanton partition function}

The method adopt in this section gives us two interesting results.
On the one hand the CFT/gauge theory computation gives us explicit answer how the classical KdV solution is deformed by $\ep_2$,
based on the underlying Virasoro algebra.
On the other hand, the prepotential $\mathcal{F}(\ep_1,\ep_2)$ of N=2 theories from localization computation is explicitly expanded as a
$q$-series, quasimodular form of the KdV charges helps us to
rewrite the expansion as large $a$-expansion with quasimodular coefficients.

\subsection{Deformed Hamiltonians of elliptic solution}

In the deformed case we do not have explicit solution to the deformed KdV equation. Therefore, in contrary to the Section 2,
now we suppose the associated CFT conformal block defines a deformation for the classical solution.
However, we do not work in the framework of CFT,
instead we compute them from the instanton partition function of deformed N=2 gauge theory. Their
equivalence is guaranteed by the widely confirmed AGT
correspondence. We propose an explicit relation between quantum Hamiltonians of elliptic solution and quasimodular expansion of the gauge theory prepotential.

Notice that in gauge theory the $\ep_2$ deformation only modestly
deforms $\mathcal{F}(\ep_1)$ as $\ep_1,\ep_2$ are symmetric
deformation, therefore we go through a procedure imitates the
classical case although there is not a spectral problem like
(\ref{lineq}) for the deformed KdV theory. When we relate classical
KdV hierarchy to gauge theory we used the fact that the monodromy
relation of KdV in equation (\ref{nulambda}) is identified with the
gauge theory relation $a=a(\tilde{\mathfrak{u}},m,q,\epsilon_1)$. In
the deformed case, gauge theory generalizes the monodromy relation
to $a=a(\tilde{\mathfrak{u}},m,q,\epsilon_1,\epsilon_2)$, which
gives the quantum monodromy of KdV solution. In order to perform
$\hbar$ expansion, we only need to slightly change some
identifications in Section 2 in order to have a smooth limit to the
classical theory. The same identification is made for the following
parameters, \be n=\f{m}{\epsilon_1},\qquad \nu=\f{2\pi
a}{\epsilon_1},\qquad
\lambda=-\f{8\pi^2\tilde{\mathfrak{u}}}{\ep_1^2}.\ee We could make
$\ep_1,\ep_2$ appear symmetrically by substituting
$\ep_1\to\sqrt{\ep_1\ep_2}$. But this is only a rescaling, it is not
essential. We also propose a relation that naturally generalizes
(\ref{ufrompoten}), \be \tilde{\mathfrak{u}}=\f{1}{2}q\f{\p}{\p
q}\mathcal{F}(\ep_1,\ep_2)+\f{(m-\ep_1)(m-\ep_2)}{24}(1-2E_2).\label{ufrompoten2}\ee
Recall that in the case of classical KdV solution we could fix the
relation (\ref{ufrompoten}) because we can independently compute the
KdV charges. In the deformed case we can only conjecture a relation. The
form of the deformed function $\mathcal{F}(\ep_1,\ep_2)$, computed from the
Nekrasov partition function, gives us a hint about the second term
on the right hand side of (\ref{ufrompoten2}), \be q\f{\p}{\p
q}\mathcal{F}(\ep_1,\ep_2)=a^2-\f{(m-\ep_1)(m-\ep_2)}{12}(1-E_2)+\mathcal{O}(a^{-2}).\ee
In the expansion coefficients of $a^{-2k}$ are
$\lambda^q_k$, $k\geqslant1$, their quasimodular polynomials have a
rigidity property, as explained below in (\ref{quantumlambda}). Our
conjectural relation maintains this property for $\lambda^q_0$.

As in the classical limit $\ep_2\to0$, the deformed prepotential $\mathcal{F}(\ep_1,\ep_2)$ should reduces
to the deformed prepotential $\mathcal{F}(\ep_1)$ in formula (\ref{potenNS}), it is reasonable to suspect the
quasimodular polynomials are preserved in someway in the $\ep_2$ deformed case. Indeed, from the Nekrasov partition function we find that for
$\lambda^q_{2k}$ in the series \be
\lambda=-\nu^2+\sum_{k=0}^{\infty}\f{\lambda_{k}^q(\hbar)}{\nu^{2k}},\label{lambdanuq}\ee
we can assemble all $q$ into $E_{2k}$, then the polynomials of
Eisenstein series appearing in the classical case are preserved,
only the coefficients $[n(n-1)]^k$ are corrected by $\hbar$. If we
define $\mathbf{N}=(n-\hbar)(n-1)$, then \be
\begin{aligned}
\lambda_0^q=&\f{\pi^2}{3}\mathbf{N}E_2,\\
\lambda_1^q=&\f{\pi^4}{36}\mathbf{N}(\mathbf{N}-\hbar)(E_2^2-E_4),\\ \lambda_2^q=&\f{\pi^6}{540}\mathbf{N}(\mathbf{N}-\hbar)(\mathbf{N}+\f{1}{2}\hbar)(5E_2^3-3E_2E_4-2E_6)-\f{\pi^6}{90}(1+\hbar)^2\mathbf{N}(\mathbf{N}-\hbar)(E_2E_4-E_6),\\
\lambda_3^q=&\f{\pi^8}{9072}\mathbf{N}(\mathbf{N}-\hbar)(\mathbf{N}^2+\f{7}{5}\mathbf{N}\hbar+\f{3}{5}\hbar^2)(35E_2^4-7E_2^2E_4-10E_4^2-18E_2E_6)\\
&-\f{\pi^8}{756}\mathbf{N}(\mathbf{N}-\hbar)[\mathbf{N}(1+\f{9}{5}\hbar+\hbar^2)+\hbar(\f{3}{2}+\f{59}{20}\hbar+\f{3}{2}\hbar^2)](7E_2^2E_4-5E_4^2-2E_2E_6)\\
&+\f{\pi^8}{126}(1+\hbar)^4\mathbf{N}(\mathbf{N}-\hbar)(E_2E_6-E_4^2),\\
\cdots\label{quantumlambda}
\end{aligned}
\ee From this perspective the polynomial ring of $E_2,E_4,E_6$ in
the expansion of $\lambda$ has certain rigidity under the second
deformation $\ep_2$.

We can inverse the relation (\ref{lambdanuq}) to get the quantum
monodromy relation that gives the quantum energy densities of the
periodic solution, \be
i\nu=\sqrt{\lambda}+\sum_{k=1}^{\infty}\f{\varepsilon^q_{k}(n,g_{2,3},\f{\zeta_1}{\omega_1},\hbar)}{(\sqrt{\lambda})^{2k-1}},\label{nulambdaq}\ee
where \be
\begin{aligned}
\varepsilon_1^q=&-\f{\pi^2}{6}\lbrace n(n-1)E_2-\hbar(n-1)E_2\rbrace,\\
\varepsilon_2^q=&-\f{\pi^4}{72}\lbrace n^2(n-1)^2E_4+\hbar[n(n-1)(E_2^2+E_4)-2n^2(n-1)E_4]\\
&+\hbar^2[n(n-1)E_4-(n-1)E_2^2]\rbrace,\\
\cdots \label{quantumcharge}
\end{aligned}
\ee $\varepsilon_k$ explicitly show how the classical charges are
deformed. This is an exact quantization, in every $\varepsilon_k$
the quantum effect grows to a finite power of $\hbar$. As we do not
observe interesting pattern in the expansion, we cease to give more
detail on this.

\subsection{Quasimodular prepotential of gauge theory}

Let us move a step backward to retrieve the prepotential by
integrating the relation (\ref{ufrompoten2}), we get an expansion $\widetilde{\mathcal{F}}(\ep_1,\ep_2)$ which is similar to $\widetilde{\mathcal{F}}(\ep_1)$ in (\ref{potenNS}).
 \beq
\widetilde{\mathcal{F}}(\ep_1,\ep_2)&=&(a^2-\f{(m-\ep_1)(m-\ep_2)}{12})\ln q+2(m-\ep_1)(m-\ep_2)\ln\eta(q)\nn\\
&\quad&-\f{m(m-\ep_1)(m-\ep_2)(m-\ep_1-\ep_2)}{48a^2}E_2+\mathcal{O}(a^{-4}).\label{modulpoten0}
\eeq
From the structure of formula (\ref{quantumlambda}), $\widetilde{\mathcal{F}}(\ep_1,\ep_2)$ contains the same quasimodular polynomials as in $\widetilde{\mathcal{F}}(\ep_1)$, only with mass terms further deformed by $\ep_2$.
The issue about the $q$-independent integration constant part is similar to the case discussed in Section 2,
they are related to the perturbative part of the  prepotential.
Among the perturbative contribution, most of terms of order $a^{-2n}$ are precisely the integration constant already introduced in $E_{2k}$,
another piece of perturbative contribution undetermined contains terms with $\ln2a$. The full deformed prepotential is
\be \mathcal{F}(\ep_1,\ep_2)=\widetilde{\mathcal{F}}(\ep_1,\ep_2)+(m-\ep_1)(m-\ep_2)\ln2a.\ee
The 1-loop and instanton contribution are related in a subtle way because in the Nekrasov partition function the perturbative part is represented as the regularized universal denominator from the instanton part \cite{instcount}, the regularized function can be computed by the logarithm of double gamma function, see e.g. \cite{no0306, bfglp}. See \ref{App} for some details.

It is instructive to write the deformed prepotential in another form, to compare with the
results derived by other methods \cite{bfglp, kpt, hkk}. We need to make a shift for the
mass, $m\to m+\f{\ep_1+\ep_2}{2}$, then the quasimodular part of the prepotential, or the terms involving $E_{2k}$, can be expanded as
 \be
\mathcal{F}_{mod}(a,m+\f{\ep_1+\ep_2}{2},q,\ep_{1,2})=\sum_{g,h=0}^{\infty}(\ep_1+\ep_2)^{2g}(\ep_1\ep_2)^h\mathbf{F}_{g,h}(a,m,E_2,E_4,E_6).\label{modulpoten0}
\ee The $\mathbf{F}_{0,0}$ coincides with the result obtained from
the Seiberg-Witten curve \cite{mnw9710}, \be
\mathbf{F}_{0,0}=-\f{m^4}{48a^2}E_2-\f{m^6}{5760a^4}(5E_2^2+E_4)-\f{m^8}{2903040a^6}(175E_2^3+84E_2E_4+11E_6)+\mathcal{O}(\f{m^{10}}{a^8}).\ee
We list a few other $\mathbf{F}_{g,h}$ up to order
$a^{-6}$ which may be useful, they are \be
\begin{aligned}
\mathbf{F}_{1,0}&=\f{m^2}{96a^2}E_2+\f{m^4}{1536a^4}(E_2^2+E_4)+\f{m^6}{414720a^6}(25E_2^3+48E_2E_4+17E_6)+\mathcal{O}(\f{m^8}{a^8}),\\
\mathbf{F}_{2,0}&=-\f{1}{768a^2}E_2-\f{m^2}{30720a^4}(5E_2^2+9E_4)-\f{m^4}{1105920a^6}(25E_2^3+84E_2E_4+101E_6)+\mathcal{O}(\f{m^6}{a^8}),\\
\mathbf{F}_{3,0}&=\f{1}{368640a^4}(5E_2^2+13E_4)+\f{m^2}{9289728a^6}(35E_2^3+168E_2E_4+355E_6)+\mathcal{O}(\f{m^4}{a^8}),\label{modulpoten1}
\end{aligned}
\ee
and
\be
\begin{aligned} \mathbf{F}_{0,1}&=-\f{m^2}{48a^2}E_2-\f{m^4}{2304a^4}(5E_2^2+E_4)-\f{m^6}{41472a^6}(11E_2^3+6E_2E_4+E_6)+\mathcal{O}(\f{m^8}{a^8}),\\
\mathbf{F}_{1,1}&=\f{1}{192a^2}E_2+\f{m^2}{23040a^4}(25E_2^2+17E_4)+\f{m^4}{276480a^6}(55E_2^3+114E_2E_4+41E_6)+\mathcal{O}(\f{m^6}{a^8}),\\
\mathbf{F}_{2,1}&=-\f{1}{184320a^4}(25E_2^2+29E_4)-\f{m^2}{7741440a^6}(385E_2^3+1386E_2E_4+1019E_6)+\mathcal{O}(\f{m^4}{a^8}),\\
&\cdots \label{modulpoten2}\end{aligned}
\ee

The quasimodular expansion of deformed prepotential has been studied by other methods, from other perspectives.
In \cite{bfglp, bfglp2} the $(\ep_1,\ep_2)$ deformed prepotential is rewritten in the quasimodular form, and a recursive relation due to the
quasimodular property of $E_2$ is found. Result for the limit case $\ep_2=0$ is related to classical CFT block, the quasimodular expansion is obtained by a WKB computation for the Lam\'{e} potential \cite{kpt, piatek}.
The failure of exact modular of $E_2$ can be remedied by introducing an antiholomorphic piece,
$N=2^*$ gauge theory is studied in the holomorphic anomaly approach in \cite{hkk}.
Here we do not have an independent new computation method because the deformation of the coefficient $[n(n-1)]^k$ is based on the computation of Nekrasov partition function, however,
we propose a precise relation to the deformation of elliptic KdV solution.

\section{Conclusion and discussion}

We present an example of relating elliptic KdV solution to
4-dimensional supersymmetric gauge theory. This can be viewed as a
natural result from a chain of very interesting connections, the
qKdV-CFT relation \cite{blz1}, the CFT/Gauge correspondence \cite{agt} and
the Gauge/Bethe correspondence \cite{ns}. The Schr\"{o}dinger
equation (\ref{lineq}) provides a key hint, the Virasoro symmetry is
the real reason that makes these connections work. Using all the
relations, we obtain the deformed prepotential of gauge theory with the UV coupling
dependence assembled into Eisenstein series, which originate from the
Weierstrass elliptic function of the KdV solution. On the other
hand, we get the quantum spectrum of elliptic KdV solution.

There is another elliptic KdV solution \cite{tv}, \be
u(x)=\sum_{j=0}^{3}n_j(n_j-1)\wp(x+\omega_j).\label{tv}\ee It is
related to the sphere 4-point conformal block, and SU(2) $N_f=4$ SYM
gauge theory. We emphasis unlike the Lam\'{e} potential, in the
potential (\ref{tv}) the modulus is not the instanton/conformal
block expansion parameter \cite{he1306}. We have confirmed that the spectrum of
the potential is related to the quasimodular expansion of the
prepotential for SU(2) $N_f=4$ SYM in the NS limit. The full Nekrasov instanton partition function would give a deformation of the spectrum.
Some formulae about integration of elliptic function, useful to compute the classical spectrum in this case, has been given in \cite{kt1305}.

It is not clear if the two examples appear just by coincidence, or
in more general case some elliptic KdV solutions are related to some
CFT blocks and gauge theories. There is a large class of CFT blocks in \cite{agt}, on
the other hand up to now the most general elliptic solution of KdV
is the Picard potential \cite{gw}. Superconformal Liouville theory is
related to gauge theory on orbifold discussed in \cite{bf1105}, while
the supersymmetric KdV is constructed in \cite{m1988}. The simplest
extended conformal algebra $W_3$ is connected to the SU(3) gauge
theories \cite{wyllard}, and on the other side is connected to the
quantum Boussinesq equation \cite{bhk2001}, and similarly they have
supersymmetric extension. For the connection of general $W$-algebra
and generalized classical nonlinear equation of KdV type, see \cite{dickey}.

\appendix

\section{Perturbative part of the prepotential\label{App}}

The perturbative prepotential includes the classical part
$\mathcal{F}^{clas}=a^2\ln q$ and the $q$-independent 1-loop part
$\mathcal{F}^{1-loop}$. The $\mathcal{F}^{1-loop}$ of the
$\ep_1,\ep_2$ deformed N=2 gauge theory can be
represented as regularized universal denominator extracted from the
instanton partition function \cite{instcount}, the function
is given by the logarithm of Barnes' double gamma function \cite{no0306}. For SU(2) $N=2^*$ theory it is
given by
\be\mathcal{F}^{1-loop}=\gamma_{\ep_1\ep_2}(2a+\ep_1)+\gamma_{\ep_1\ep_2}(2a+\ep_2)-\gamma_{\ep_1\ep_2}(2a+m)-\gamma_{\ep_1\ep_2}(2a-m+\ep_1+\ep_2).
\ee The following integral form is useful for computing the large
$x$ asymptotic expansion, \be
\gamma_{\ep_1\ep_2}(x)=\lim_{s\to0}\f{d}{ds}\f{1}{\Gamma(s)}\int_0^{\infty}dtt^{s-1}\f{e^{-tx}}{(1-e^{-\ep_1t})(1-e^{-\ep_2t})},\qquad
|\text{Arg}(x)|<\pi.\ee

We can compute the asymptotic expansion in the semiclassical region
$a\gg m$. The term containing logarithm is simply
$(m-\ep_1)(m-\ep_2)\ln2a$. The remaining terms are of the form
$a^{-2n}f(m,\ep_{1,2}), n\geqslant1$, if we shift the mass by $m\to
m+\f{\ep_1+\ep_2}{2}$ then it expands as \be
\sum_{g,h=0}^{\infty}(\ep_1+\ep_2)^{2g}(\ep_1\ep_2)^h\mathbf{f}_{g,h}(a,m).\ee
We indeed find $\mathbf{f}_{g,h}=\mathbf{F}_{g,h}|_{q=0}$.

\section*{Acknowledgments}

I would like to thank  Andrei Mikhailov for valuable discussion and reading the draft.
I also thank the referee for detailed and very helpful comments.
This work is supported by the FAPESP No. 2011/21812-8, through IFT-UNESP.

\end{document}